\documentclass{aa}
\usepackage[varg]{txfonts}
\usepackage{amsmath}
\usepackage{amssymb}
\usepackage{mathrsfs}
\usepackage{natbib}
\usepackage{aas_macros}
\usepackage{gensymb}
\usepackage{appendix}
\usepackage{multirow}
\usepackage[font=small,labelfont=bf]{caption}

\newcommand{\refrevision}[1]{#1}

\usepackage{graphicx}

\newcommand\prox{Proxima Centauri}
\newcommand\rdwarf{M-dwarf}

\newcommand{\ha}{H$\alpha$}

\newcommand{\harps}{{\sc harps}}
\newcommand{\asas}{{\sc asas}}
\newcommand{\uves}{{\sc uves}}

\newcommand{\hst}{{\sc hst}}

\newcommand{\scipy}{{\sc scipy}}
\newcommand{\astroml}{{\sc astroml}}
\newcommand{\gatspy}{{\sc gatspy}}
\newcommand{\matplot}{{\sc matplotlib}}

\newcommand{\horn}{sub-peak}

\newcommand\Notnow[1]{}

\newcommand\IfThesis[1]{}

\newcommand\paperorthesis{paper}
\newcommand\FirstP{We}
\newcommand\Firstp{we}
\newcommand\Firstobj{us}
\newcommand\Firstposs{our}
\newcommand\FirstPoss{Our}

\begin{document}

\title{Calculations of periodicity from {\ha} profiles of {\prox}}

\author{John M. Collins\inst{\ref{uherts}}\and Hugh R.A. Jones\inst{\ref{uherts}}\and John R. Barnes\inst{\ref{openu}}}

\institute{University of Hertfordshire, College Lane, Hatfield, Herts, AL10 9AB, UK\label{uherts} \and 
Department of Physical Sciences, The Open University, Walton Hall, Milton Keynes, MK7 6AA, UK\label{openu}}

\date{\today}

\abstract{{\FirstP} investigate retrieval of the stellar rotation signal for {\prox}. {\FirstP} make use of
  high-resolution spectra taken with {\uves} and {\harps} of {\prox} over a 13-year period as well as photometric
  observations of {\prox} from {\asas} and {\hst}. {\FirstP} measure the {\ha} equivalent width and {\ha} index,
  skewness and kurtosis and introduce a method that investigates the symmetry of the line, the Peak Ratio, which appears
  to return better results than the other measurements. {\FirstPoss} investigations return a most significant period of
  \refrevision{82.6} $\pm$ \refrevision{0.1} days, confirming \refrevision{earlier} photometric results and ruling out a
  more recent result of 116.6 days which \refrevision{{\Firstp} conclude to be an alias induced by the specific {\harps}
  observation times}. {\FirstP} conclude that whilst spectroscopic {\ha} measurements can be used for period recovery,
  in the case of {\prox} the available photometric measurements are more reliable. {\FirstP} make 2D models of {\prox}
  to generate simulated {\ha}, finding that reasonable distributions of plage and chromospheric features are able to
  reproduce the equivalent width variations in observed data and recover the rotation period, including after the
  addition of simulated noise and flares. \refrevision{However the 2D models used fail to generate the observed variety
  of line shapes measured by the peak ratio. {\FirstP} conclude that only 3D models which incorporate vertical motions
  in the chromosphere can achieve this.}}


\maketitle

\protect\label{firstpage}

\section{Introduction}
\protect\label{section:intro}

{\rdwarf} stars account for over 75\% of the stars within 25 pc of the Sun \citep{winters15}; and indeed, our nearest
neighbour, Proxima Centauri, is an M5.5V star. Despite the prevalence of {\rdwarf}s, many aspects of their activity have
remained less well characterised than more massive stars, mainly because of their inherent faintness. Since stars become
fully convective at around M4V, the nature of magnetic activity and the relationship to the rotation period in the later
\rdwarf s has been of particular interest, for example in \citet{mohanty03} and \citet{reiners08}.

An understanding of the origin of periodic signals arising in a stellar system is important in the identification of
exoplanets. For example, differing conclusions as to whether reported planets have been validly detected by their period
have been offered, for example for GJ581 (\citealt{robertson14,robertson14a,tuomi13aug,hatzes15}).  The behaviour of the
{\ha} line is a potentially important diagnostic because it is sensitive to magnetic activity and is a strong line
usually seen in emission in later {\rdwarf} stars.

\protect\label{section:intro_proxcen} At a distance of only 1.3 pc, Proxima Centauri is a bright M5.5V star with a
magnitude of 11.13 in the V-band. The rotation period of Proxima is nevertheless uncertain. \refrevision{The rotation
  period is of interest for various studies, including flare cycles \citep{davenport16} and for the correct
  identification of radial velocity signals from orbiting planets \citep{angladaescude16} and subsequent work
  \citep{ribas16}.}  Previous studies have reported periods ranging from 31.5 $\pm$ 1.5 days \citep{guinan96}, through
41.3 days \citep{benedict93} to between 82 and 84 days \citep{benedict92,benedict98}. \citet{kurster99} found that the
period is not less than 50 days \citep{kurster99}, while a more recent value of 82.5 days \citep{kiraga07} has confirmed
earlier estimates. All those measurements were obtained by photometry. \refrevision{An alternative method for establishing
periodicity is the use of \ha, e.g., \citet{feinstein76}.} An even longer rotation period of 116.6 days
\citep{suarezmascareno15} has been suggested from spectroscopy, in particular via a study of the {\ha} line
\refrevision{as the {\ha} Index measure} from the {\harps} data used in this {\paperorthesis}. \citet{cincunegui07}
reported a 442-day activity cycle, \refrevision{by consideration of the FWHM\footnote{This is functionally identical to
    Equivalent Width} of the {\ha} line taken from observations using the 2.15m telescope of CASLEO}.

Here {\Firstp} investigate whether periodicity can be identified in the morphology of the {\ha} line in high-resolution
spectra such as those obtained from the Ultraviolet and Visual Echelle Spectrograph ({\uves}) at the 8.2m Very Large
Telescope (VLT, UT 2) and the High Accuracy Radial velocity Planet Searcher ({\harps}) at the ESO La Silla 3.6m
telescope. {\prox} also shows frequent flaring activity and the presence of these flares is useful for determination of
whether, and to what extent, they affect estimates of the rotation period.

\section{Periodicity of {\prox} from photometric measurements}
\protect\label{section:asas}

To process the data for all the periodicity studies in this \paperorthesis, {\Firstp} used a variety of Lomb-Scargle
routines. \refrevision{This was necessary as different implementations return different periods, especially when peak
  significance is low.} The Lomb-Scargle routine in \textit{Numerical Recipes}\footnote{This can be
  obtained from http://numerical.recipes/.}, modified to return FAP values for all peaks, is valuable for the
periodograms where there are comparatively clear-cut peaks, however for many of the spectrographic results it is unable
to return any clear periods. For these cases software was written in Python using alternative Lomb-Scargle routines
provided by {\scipy} library \citep{jones01}, the {\astroml} library, \citep{vanderplas12} and the {\gatspy} library,
\citep{vanderplas15}, comparing the results. {\FirstP} did this because in some cases the results widely differed and it
gave an assessment of the stability of the calculations. In most cases the {\gatspy} routine appeared to be the most
reliable \refrevision{and was used where the \textit{Numerical Recipes} routine failed to find the periods sought. This
  is consistent with the study by Jake Vanderplas\footnote{See
    https://jakevdp.github.io/blog/2015/06/13/lomb-scargle-in-python/}.}

As nearly all previous measurements of periodicity in {\prox} were made using photometric observations, in this
{\paperorthesis}, {\Firstp} first present results obtained from the photometric observations for {\prox} taken from the
V-Band (there were no data for the I-Band) of the All Sky Automated Survey (\asas), \citep{pojmanski97}, which contains
data between the periods December 2000 to September 2009. \refrevision{This was done to experiment with the binning and
  also for the evaluation of the FAP and error bars from the {\harps} data, as described in Section
  \ref{section:asasfap}.}

\begin{figure}[!htbp]
\begin{center}
\vspace{-.25cm}
\includegraphics[scale=0.25]{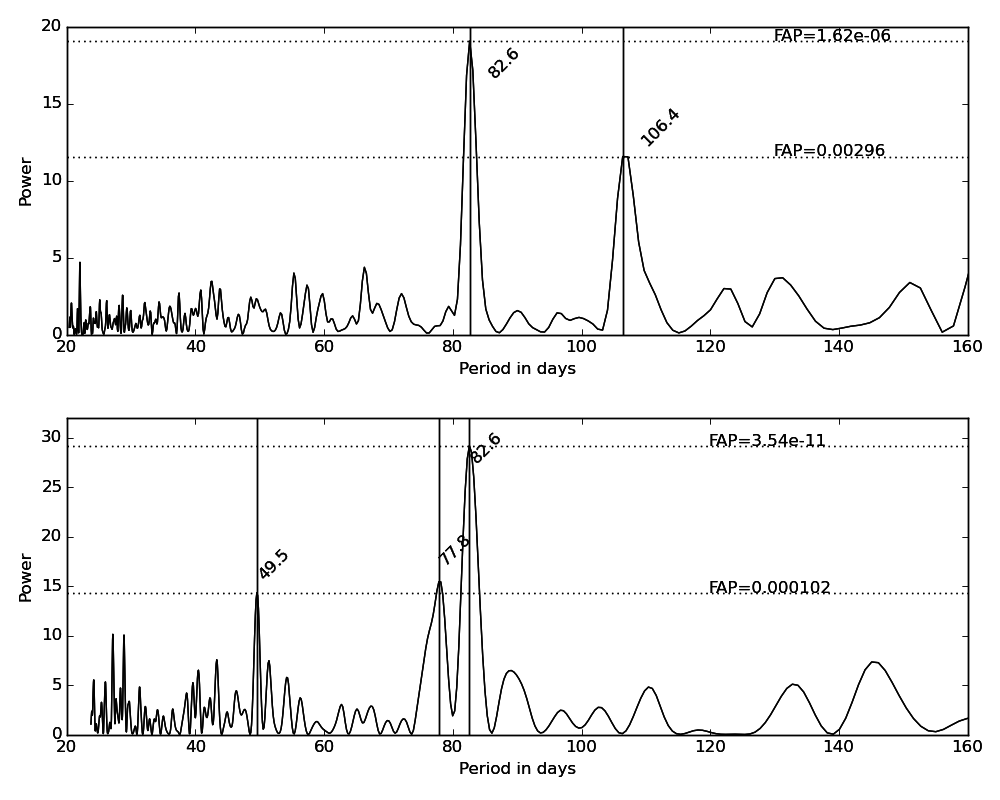} \\
\vspace{-.5cm}
\end{center}   
\caption{The upper panel is a periodogram from the {\asas} database for {\prox} second aperture, binned to 1 day.  The
  lower panel is a similar periodogram from the {\hst} data discussed in \citet{benedict98}. FAP values for the
  strongest peaks are shown as the dotted lines.}
\protect\label{fig:asasexample}
\end{figure}

As indicated by the {\asas} guidelines\footnote{http://www.astrouw.edu.pl/asas/explanations.html}, with {\prox} set out
in the {\asas} data as having magnitude 11 in the V-band, {\Firstp} took the data from the second aperture. {\FirstP}
are only considering the ``best'' (grade A) data from this aperture, which has 970 points. As some of the observations
were overlapping in time, {\Firstp} binned these to 1 day, which reduced the number of points to 624. {\FirstP} then
obtained the periodogram shown in the upper panel of Fig. \ref{fig:asasexample}. Periods between 20 and 160 days were
taken in this case. It is noticeable that there only two significant peaks, at 82.6 and 106.8 days, with negligible
FAPs.

{\FirstP} also obtained a periodogram from the {\hst} data discussed in \citealt{benedict92,benedict98} consisting of
171 points obtained between July 1995 and January 1998, later enhanced so the last 18 points extended to January 2001,
obtaining the lower panel of Fig. \ref{fig:asasexample}, again taking between 20 and 160 days. The observation times
were either on separate days, or spaced out evenly throughout a single day with the result that binning this data would
have reduced the data unacceptably, so {\Firstp} did not bin the {\hst} data.

It is clear that there is consistent agreement between these results with a strong period of 82.6 $\pm$ 0.1 days and in
agreement with the rotation period given in \citet{benedict98} and confirmed in \citet{kiraga07}.  The {\asas} data also
includes a reasonably strong additional signal of 106.5 $\pm$ 0.2 days. Taking account of the possibility of this being
associated with some interaction between the main period and some other
period, % as discussed in Section \ref{section:multiperiod},
{\Firstp} considered whether this might be a ``beat'' period between the observation years and the rotational period, as
we noticed the relationship $ \frac{1}{\frac{1}{82.6} - \frac{1}{365.25}} = 106.7 $ and {\Firstp} also noticed that $
\frac{1}{\frac{1}{82.6} + \frac{1}{365.25}} = 67.4 $, which appears as the third peak observed in some of the
periodograms obtained from all the {\asas} apertures. \refrevision{{\FirstP} considered the window functions of the
  {\asas} and {\hst} data but were unable to find any significant periods around those values.}

{\FirstP} also searched for very long periods up to the period spanned by the data in each case, however {\Firstp} did
not find any strong periods, in particular nothing close to the 442 days reported in \citet{cincunegui07}
\refrevision{based upon FWHM of {\ha} peaks from observations using CASLEO}.

In any event the {\asas} and {\hst} provides a convenient benchmark for assessing the accuracy and reliability of the
other methods based on the {\ha} line.

\section{Spectra of \prox}
\protect\label{section:proxima}

{\FirstP} looked at two sources of spectra for \prox, the {\uves} spectra taken between 10th and 14th March 2009 studied
in \citet{fuhrmeister11} and the {\harps} spectra, with 260 data points between May 2004 and January 2014 from the ESO
archive. The {\uves} data were obtained with a 0.8/1.10'' slit, yielding a resolution of approximately 60,000, while the
resolution of {\harps} is approximately 120,000.  \refrevision{The spectral range of {\uves} used for the observations
  is 6380 to 10250 $\AA$ while the fixed format of {\harps} gives wavelength coverage from 3780 to 6910 $\AA$.}
{\FirstP} also studied the X-ray data from XMM-Newton used in the \citet{fuhrmeister11} paper (Provided by Fuhrmeister,
priv. comm.) to identify any association between strong X-ray values and possible corresponding changes to the {\ha}
profile. \refrevision{All the observation times of individual spectra were adjusted for the barycentric dates and
  spectra for corresponding radial velocities.}

\subsection{{\harps} and {\uves} spectra of \prox}

As mentioned in \citet{mohanty03} and subsequent papers such as \citet{jenkins09} and \citet{barnes14} the spectra of
the later of late \rdwarf s from approximately M5 onward, usually show {\ha} in emission, Several of the \rdwarf s
illustrated in \citet[Fig. 6]{barnes14} additionally show a distinct ``horned'' appearance, due to a certain amount of
self-absorption affecting the centre of the {\ha} peak. {\prox} consistently shows this pattern, which is displayed in
\citet[Fig. 14]{fuhrmeister11}. The two \horn s surround a local minimum. As well as the equivalent width of the entire
{\ha} peak, the two \horn s vary in relative size over time on either side of the local minimum, which does not greatly
change in morphology over time. This would appear to be because a more symmetrically-distributed spectral line from the
photosphere is overlaid with plage and chromospheric effects which are asymmetric or localised to regions.
% These may be compared with the H, K and Ca II lines discussed in \citet{rauscher06}.
In this {\paperorthesis} {\Firstp} seek to study the variations in the line and \horn s to see if periodicity may be reliably recovered.

\subsection{{\ha} Line measurements}
\protect\label{section:linemeas}

\begin{figure}[!htbp]
\begin{center}
\includegraphics[scale=0.40]{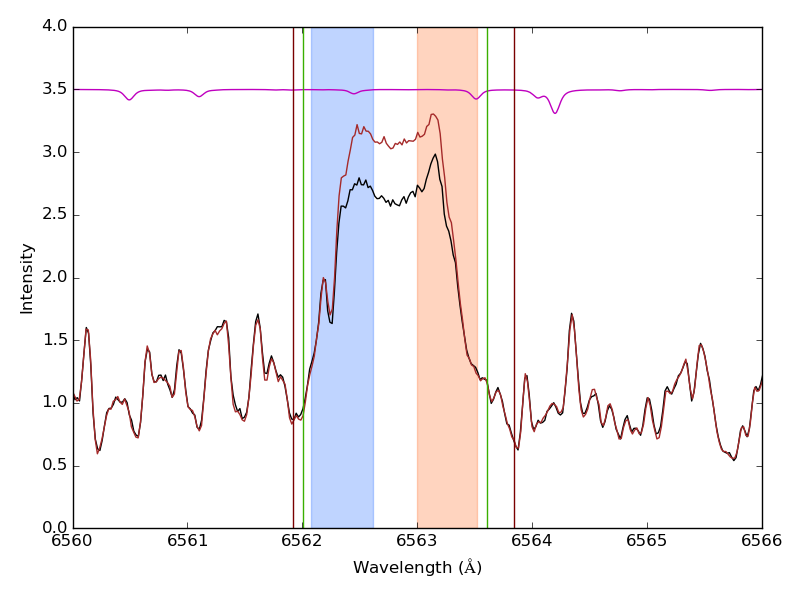} \\
\vspace{-.5cm}
\end{center}   
\caption{The {\ha} region of example spectra of {\prox} taken from {\harps} on 27 May 2004 02:10:14 UTC (black) and 15
  March 2006 09:16:35 (brown). The region delineated with the dark red solid vertical lines shows the region used for
  calculation of the {\ha} equivalent width in this \paperorthesis. The regions shaded in red and blue respectively show
  the regions used for calculation of the sizes of the two \horn s. \refrevision{The vertical
purple lines mark the region chosen for calculation of the Equivalent Width and
the vertical green lines mark the region chosen in \citet{suarezmascareno15} for calculation of the {\ha} Index.}}
\protect\label{fig:harpsfirstha}
\vspace{-.4cm}
\end{figure}

In Fig. \ref{fig:harpsfirstha}, {\Firstp} show two example spectra, in this case from {\harps} nearly
2 years apart, clearly showing the changes in the amplitude and shape of the {\ha} line. Fig. \ref{fig:harpsfirstha}
also illustrates the regions used to investigate periodic variability. {\FirstP} used the {\harps} data referred to in
\citet[Table 3]{suarezmascareno15}, which consists of 260 spectra taken between 27 May 2004 and 14 Jan 2014.
%together wish some additional data collected between 19 Jan 2016 and 30 Mar 2016.
To calculate the equivalent width, the spectra were normalised by iteratively fitting a cubic polynomial to all the
points in all the spectra, apart from the {\ha} region, then excluding points outside 2 standard deviations above or
below the fitted polynomial to eliminate both emission and absorption lines. With the normalised spectra, {\Firstp}
computed the equivalent widths and what {\Firstp} called the ``peak ratio'', defined as the ratio of the mean values of
the two \horn s. The ratio calculated is the mean value of the ``red'' \horn, i.e. that for the longer wavelength,
divided by the mean value of the ``blue'' {\horn} (i.e. the ratio of the longer wavelength to the shorter wavelength
peak). For calculation of the Equivalent Width, since pixel-wavelength scales are not identical, values of the flux are
interpolated up to the boundaries of the regions chosen, to minimise integer pixel noise effects.
% The number of pixels in the regions, as highlighted in Fig. \ref{fig:harpsfirstha} are either 35 or 36 for the
%{\ha} region and 11 or 12 for the red {\horn} and 14 or 15 for the blue {\horn} on {\uves}. The corresponding numbers on
%{\harps} are 98 or 99, 26 or 27 and 27 or 28 respectively.

{\FirstP} restricted the {\ha} region for calculation of the Equivalent Width to the minima on either side of the peak
to that from 6561.917{\AA} to 6563.839\AA) as delineated by the dark red vertical lines. The regions selected for the
blue and red \horn s are shaded in blue and red and run from 6562.072{\AA} to 6562.613{\AA} and 6563.000{\AA} to
6563.517{\AA} respectively. These regions were chosen to optimise variability in the line profiles \refrevision{to give
  the highest degree of variability with the smallest amount of noise}.

Note that the regions selected for calculation of the Peak Ratios are not quite the same width, the ``blue'' {\horn}
region having a width of 0.541{\AA} and the ``red'' {\horn} region a width of 0.517\AA. This is because in the observed
data the ``red'' {\horn} tends to be higher but narrower than the ``blue'' \horn. As the Peak Ratio is the ratio of the
mean value in the two areas, this should not be of significance.

At the top of Fig. \ref{fig:harpsfirstha} is displayed the telluric line spectrum for an air mass of 1.4, to which 2.5
has been added for clarity of display. As demonstrated in \citet[fig. 1]{reiners15}, the telluric effects are negligible
in this region. (The Gaussian used to simulate {\ha} in that paper is considerably broader than that observed in \prox,
so the telluric line at 6564.2{\AA} can impinge on the former.) All the spectral lines identifiable from the Vienna
Atomic Line Database (VALD) are TiO transitions, with the exception of a MgH line at 6564.29\AA.

In \citet{suarezmascareno15}, the authors use an \textit{{\ha} index}, based in turn upon the work in
\citet{gomesdasilva11}, computed by the formula $ H\alpha_{index} = \frac{H\alpha_{core}}{H\alpha_L + H\alpha_R} $ in
which $ H\alpha_{core} $ is defined as the bandpass of width 1.6{\AA} centred on 6562.808{\AA} and $ H\alpha_L $ and
$H \alpha_R $ are defined respectively as continuum bands of widths 10.75{\AA} and 8.75{\AA} centred on 6550.87{\AA} and
6580.31\AA. An important difference between this and calculation of the Equivalent Widths and Peak Ratios is that the
spectra do not have to be normalised prior to the {\ha} Index calculation.

The region chosen for calculation of the {\ha} Equivalent Width in this {\paperorthesis} is slightly wider than that
chosen for the {\ha} index in the \citet{suarezmascareno15}. This was chosen as in {\Firstposs} view it on average
encompassed the base of the {\ha} peak more accurately.  In practice there was negligible difference between the
calculated results for either method using the two pairs of limits or adjusting the continuum regions.

\begin{figure}[!htbp]
\begin{center}
\vspace{-.25cm}
\includegraphics[scale=0.4]{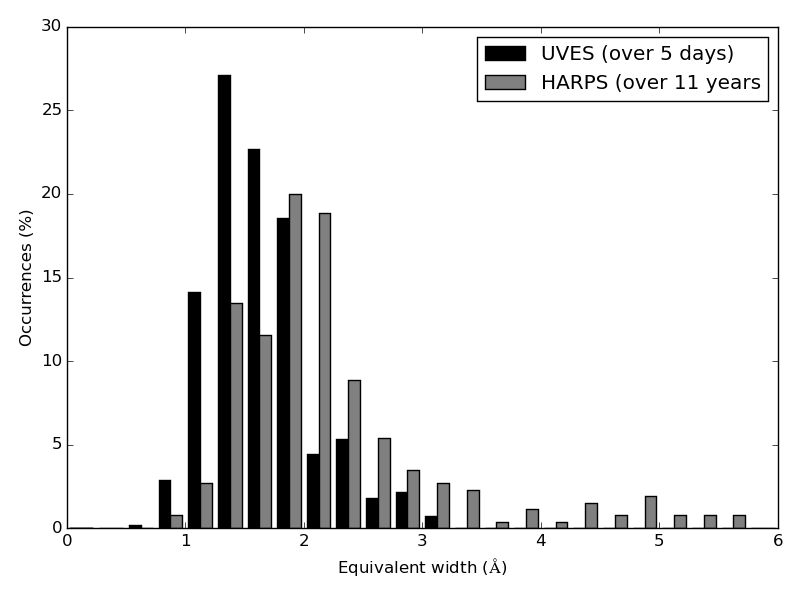} \\
\vspace{-.5cm}
\end{center}
\caption{Histograms of equivalent widths for {\uves} in blue and {\harps} in green, expressed as percentages with the
  same X axis scale. All the {\uves} spectra results are shown apart from those for one which appeared to be just noise
  (12 March 2009 UTC 02:31:11). {\harps} spectra are omitted for four outlying cases which appeared to be dominated by
  flares at 16 July 2004 UTC 01:52:40, 27 March 2011 UTC 05:20:09, 05 May 2013 UTC 03:31:16 and 5 May 2013 UTC 03:41:47
  with values of 21.69, 18.12, 6.76 and 6.19 respectively.
%  {\FirstP} also noted that these spectra had a noticeable
%  He-6678 line associated with these which appears at first analysis to disappear rapidly after the peaks, not appearing
%  significantly on higher {\ha} Equivalent Width values outside those four spectra, suggesting that those four might be
%  major flares.
}
\protect\label{fig:proxhists}
\vspace{-.25cm}
\end{figure}

Histograms of the Equivalent Widths are shown in Fig. \ref{fig:proxhists}. Note that all the Equivalent Widths from the
{\uves} data are displayed, but the four very highest from the {\harps} data are omitted, which have Equivalent Widths
of over 6, listed in the caption to Fig. \ref{fig:proxhists}.

% Done with 25 bins scale 0:6

{\FirstP} calculated the Equivalent Widths, {\ha} Index, Peak Ratios, Skewness, Kurtosis \footnote{\refrevision{The last
    two were calculated using the \textit{Scipy} statistical routines from the {\ha} region used in the
    calculation of the Equivalent Width.}} from the {\harps} data for {\prox}. Also calculated were Equivalent Widths
and Peak Ratios from and residual {\ha} lines, created by division of each spectrum by the mean of the 5 spectra with
the lowest equivalent widths. The median value and standard deviation of the {\harps} Equivalent Width was 2.0 $ \pm $
1.8 and these values are used the {\Firstposs} calculation of periodicity in Section \ref{section:harps}. {\ha} Index
values were very similar to the Equivalent Widths in all cases.

\subsection{Flares on {\uves} data and X-ray values}
\protect\label{section:uvesflares}

In \citet[fig. 1 to fig. 3]{fuhrmeister11} the measured flux for various wavelengths for each of the three observation
nights are presented. It should be noted that the X-ray flux is much greater on the third day and the scale is much
smaller in the third figure of that paper. The {\uves} data showed a large flare during the third of the observation
periods starting at approximately 06:15 on 14th March 2009. Both the equivalent width and X-ray counts rapidly reached a
peak, with the equivalent width peaking approximately a minute before the X-ray count peaked. The equivalent width
reached a similar level at the end of the first observation period to that which it reached during the flare in the
third, albeit much more slowly, but with only very slight evidence of a corresponding increase in the X-ray
count. However there was an increase in the {\uves} optical''blue'' flux on the first day, as shown in
\citet[fig. 1]{fuhrmeister11} corresponding to the higher Equivalent Widths suggestive of a flare.

There is no corresponding X-ray data available for the {\harps} data, but the {\uves} data suggests that {\ha}
Equivalent Width increases with flares. {\FirstP} thus selected the higher values of Equivalent Width in the {\harps}
data as indicative of flares. After some experimentation with investigation of periodicity, the effects of possible
flares seemed to be minimised without losing too much data if the proportion of data with the lower 90\% of Equivalent
Widths were selected. In both {\uves} and {\harps} this was approximately one standard deviation from the median, 3.8 in
the case of {\harps}.

\subsection{Recovery of periods from {\harps} data}
\protect\label{section:harps}

\begin{figure}[!htbp]
\begin{center}
\includegraphics[scale=0.20]{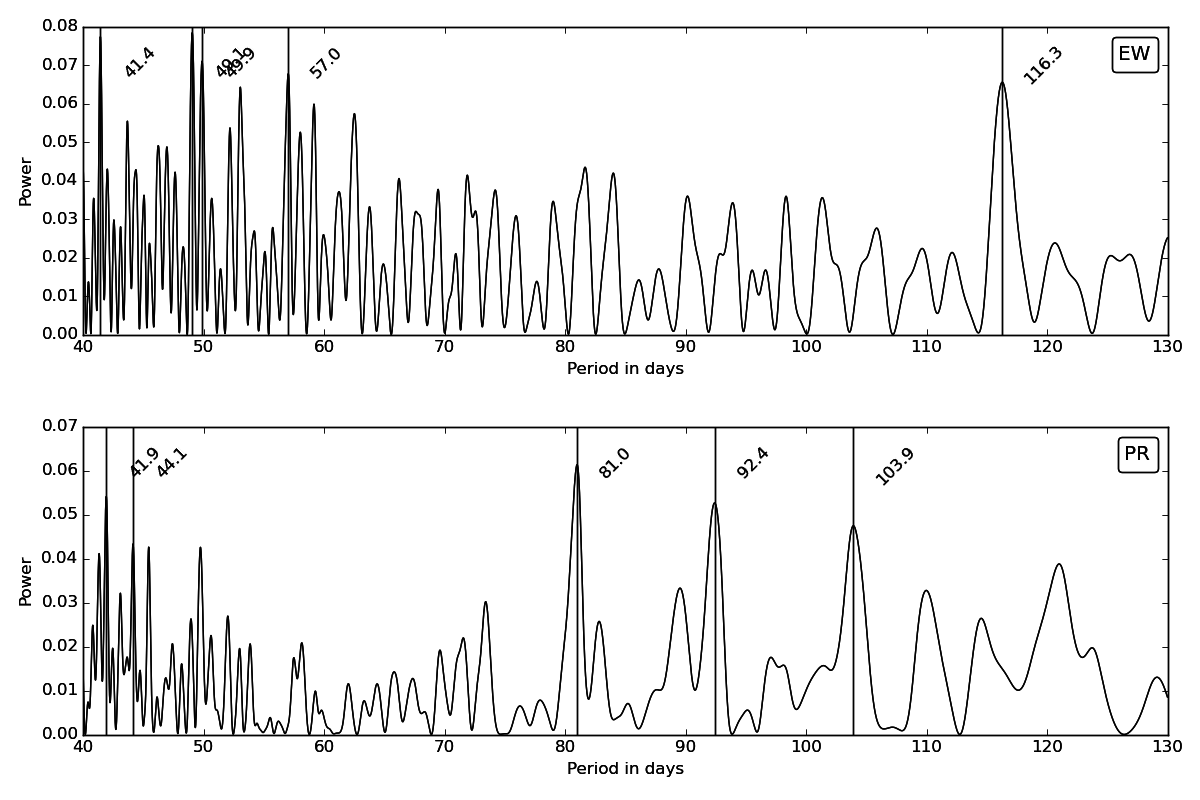} \\
\vspace{-.5cm}
\end{center}   
\caption{This figure shows sample periodograms from the {\ha} peak of the {\harps} data, for the range between 40 days
  and 130 days in steps of 0.01 days. The top panel shows the periodogram derived from the Equivalent Widths (EW) and
  the bottom panels ones derived from the Peak Ratios (PR). None have any clipping or binning of the data.  The
  strongest five peaks are marked on the vertical black lines.}
\protect\label{fig:harpspgrams1}
\end{figure}

{\FirstP} computed periodograms for periods between 40 days and 130 days, in steps of 0.01 days (14 minutes, 20 seconds)
taking into account the minimum period of 50 days given by \citet{kurster99} and the 82 days of
\citealt{benedict92,benedict93,benedict98,kiraga07} and the 116.6 days of \citet[Table 3]{suarezmascareno15}. Two sample
periodograms are exhibited in Fig. \ref{fig:harpspgrams1}.  {\FirstP} tried all the methods described above in Section
\ref{section:proxima} and various combinations thereof. All the results were obtained using all three Python
Lomb-Scargle routines as it was discovered that the results from these varied widely. It was difficult to identify any
of these periods using the \textit{Numerical Recipe}s Lomb-Scargle routine, it either failed to find them, or it
reported an FAP of 1.0.

Results from Equivalent Width calculations and {\ha} Index calculations were almost completely identical in all
cases. {\FirstP} were able to reproduce, not necessarily as the strongest peak in the periodograms, the 116.6 days of
\citet{suarezmascareno15}, although a period of 116.3 days was obtained, using Equivalent Width, {\ha} Index, Skewness
and Kurtosis measurements on untransformed data. However this disappeared as soon as any clipping or binning of the data
was performed.

\refrevision{There did not appear to be any consistent way of improving the performance of the various methods of
  measurement by treatments of the data. Treatments which were tried included:}
\refrevision{\begin{itemize}
\item Clipping data with extremes of equivalent width.
\item Binning to various periods ranging from 30 minutes through to 7 days.
\item Restricting the dataset to subsets within periods from 6 months through to 2 years, in case differing activity levels over the period of the whole set
  were affecting the results.
\item Taking of residual spectra by dividing each spectrum by the mean of various-sized selections of the spectra with
  lowest equivalent widths\footnote{Very little difference was observed in the size of selections used and five was
  adopted in the end}.
\end{itemize}}

\refrevision{This yielded approximately 100 different treatments of the data. Each of these were processed with each Lomb-Scargle
routine available.}

\refrevision{For each measurement technique, {\Firstp} assessed the performance by considering how approximately how often it
delivered:}

\refrevision{\begin{itemize}
\item The period of 82.6 days as the highest peak in the periodogram.
\item The period of 82.6 days or the obvious alias of half this period of 41.3 days as the highest peak.
\item The period of 82.6 days as one of the top 5 peaks in the periodogram.
\item The period of 82.6 days or 41.3 days as one of the top 5 peaks in the periodogram.
\end{itemize}}

For Equivalent Widths the period of 82.6 days never appeared as the strongest peak, \refrevision{in 14\% of the results}
as one of the top five peaks and 82.6 days or 41.3 days as one of the top five peaks in 43\% \refrevision{of the results}.

For Peak Ratios the period of 82.6 days appeared \refrevision{in 29\% of the results} as the strongest peak,
\refrevision{in 48\% of the results} as one of the top five peaks and 82.6 days or 41.3 days as one of the top five
peaks \refrevision{in 62\% of the results}.

\refrevision{Skewness and Kurtosis measurements were intermediate in performance between these extremes but were much
  less affected by variations in the treatments of the data such as clipping, binning or restriction to subsets by date.}

Again as with the {\asas} and {\hst} data, {\Firstp} checked for other periodic signals of up to the span of the data,
but were unable to discern any strong period and in particular no sign of the 442-day period reported in
\citet{cincunegui07} \refrevision{(using measurements of the FWHM of the {\ha} line)}.

\subsection{Comparison of {\asas} and {\harps} for period recovery}
\protect\label{section:asasfap}

{\FirstP} chose to look in more detail at the {\asas} data which offers similar sampling to the {\harps} data discussed
in Section \ref{section:harps} above. Of particular importance is the False Alarm Probability of periods recovered from
the spectroscopic data as well as the error bar from the calculated periods. None of the three Python routines directly
return a False Alarm Probability and the \textit{Numerical Recipes} routine always returned an FAP of 1 if the periods
were found at all, so {\Firstp} devised a Monte Carlo method of estimating this and at the same time estimating the
uncertainty on the period recovered from the {\asas} results.

The {\asas} data has many more observation times than the spectroscopic data, with 970 points for each aperture, which
even after binning to one day, reduces to 624 points. In contrast to this, the {\harps} data studied in Section
\ref{section:harps} has 260 spectra, which after clipping to less than 1 standard deviation above the median and binning
as described in Section \ref{section:uvesflares} reduces to 55 points.

\begin{figure}[!htbp]
\begin{center}
\vspace{-.25cm}
\includegraphics[scale=0.40]{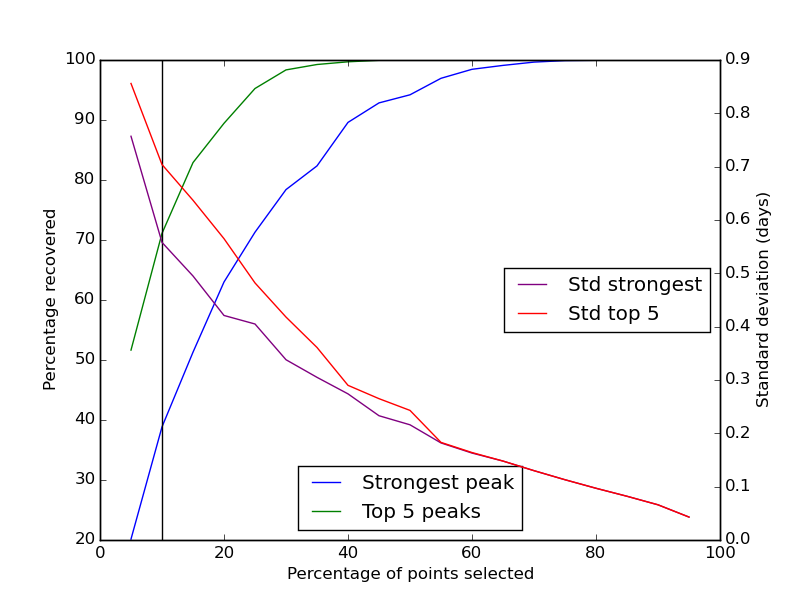} \\
\vspace{-.5cm}
\end{center}
\caption{In this figure is illustrated the effects of randomly selecting a given proportion of the {\asas} data in terms
  of whether the same period of 82.6 days is recovered and the error in this result. The black vertical line represents
  the proportion of the {\asas} data which corresponds to the number of spectra in the clipped and binned
  \refrevision{to one day} {\harps} data so that the relative performances of the spectroscopic results can be
  compared.}
\protect\label{fig:asasprop}
\vspace{-.5cm}
\end{figure}

To study how the performance of the recovery of the period is affected by the reduction in the data, {\Firstp} assumed
for {\Firstposs} purposes that the 82.6 day period is correct and tested how the recovery of this period is affected by
random selection of subsets. First {\Firstp} took the {\asas} data after binning to one day and then took various
percentage-sized randomly-selected subsets of this data, recalculating the periods, noting whether a value close to the
correct period was returned as the strongest peak, within the 5 strongest peaks, or not at all. If the period was
recovered, {\Firstp} recorded the RMS error, i.e. difference from 82.6 days. {\FirstP} took sizes of subset between 5\%
and 95\% in steps of 5\%. For each percentage sized subset, the process was repeated 2,000 times. The results are
illustrated in Fig. \ref{fig:asasprop}.

In Fig. \ref{fig:asasprop} are shown 4 results. On the X-axis is shown the percentage sized subset of the binned {\asas}
data which was used. On the left Y-axis is shown the percentage of recovery (i.e. 100\% minus the FAP) of the correct
period. On the right Y-axis is shown the RMS error in the correctly-recovered period. The blue line shows the percentage
recovery of the correct period as the strongest peak with various percentage sized subsets of the data. The green line
shows the percentage recovery of the correct period as one of the five strongest peaks, not necessarily the strongest
peak. It can be seen that the former reaches 100\% recovery at about 70\% and the latter at about 45\%. The other two
lines show the RMS error in the correctly-returned results as various sized subsets of the data are selected. The purple
line shows the RMS error in the results corresponding to the case where only the strongest peak is selected and the red
line that where the correct result is found in one of the top five peaks. It is noticeable that this reaches less than
0.1 days in both cases, lending weight to the conclusion that the uncertainty in the 82.6 day peak found by {\asas} and
{\hst} is of the order of 0.1 days.

Finally {\Firstp} marked in, as the vertical black line in Fig. \ref{fig:asasprop}, the percentage sized subset of the
data which corresponds to the number in the clipped and binned {\harps} data \refrevision{(the binning was to one day
  also)} so that a comparison can be made with the performance that against the {\asas} data reduced to the same number
of results. It can be seen that this intersects the blue line, representing the period found as the strongest peak, at
just under 40\% and the green line, where the period is found as one of the top five peaks, at just under 70\%. As seen
in Section \ref{section:harps} the corresponding figures for Peak Ratios on {\harps} are 29\% and 48\% respectively and
for Equivalent Widths 0\% and 14\%. With the same number of data points, therefore, the results from {\ha} Peak Ratios
are roughly half as good as the photometric results and the Equivalent Width results are roughly a quarter as good as
those for the Peak Ratios.

\section{Modelling of {\prox} spectra}
\protect\label{section:modelling}

In order to develop and refine the methods for evaluation of the periodicity of the sub-peaks in the {\prox} spectra,
{\Firstp} used a version of the ``Doppler Tomography of Stars'' (DoTS) modelling software
\citep{CCamerondotsa}. Although DoTS was written to recover surface imhomogeneities from time series spectra, here
{\Firstp} use the forward modelling routines to generate synthetic spectra, with some modifications. Specifically,
{\Firstp} construct a 3D model of the star, covered in a finite number of pixels. The intensity of each pixel can vary
from a photospheric value to a value appropriate for plage. In order to obtain the appropriate photospheric intensity
for each pixel at a given rotation phase of the 3D stellar model, {\Firstp} used the 4-parameter limb darkening law
introduced by Claret from Phoenix model atmospheres \citep{claret00a} for an effective temperature of 3000K. The plage
intensities were calculated according to \citet[Section 4.1]{unruh99}, who identified the centre to limb variability
from plage regions relative to the photospheric (quiet) intensity for the Sun. Since no such observations exist for
other stars, {\Firstp} adopted the same law with appropriate facular contrasts for {\ha} wavelengths (see \citet[figs 3
\& 4]{unruh99}).

Since {\Firstp} wish to simulate the {\ha} line profile, a local intensity profile is assumed for the photosphere and
the plage. For inactive photospheres of \rdwarf s of a similar spectral type to {\prox}, {\ha} is not visible (e.g. see
{\ha} profile in \citet[fig. 6]{barnes14} for GJ1061). Hence for the quiet photosphere, {\Firstp} assume a flat
continuum. For active stars, {\ha} possesses a characteristic emission profile with self-absorption, resulting in a
double-peaked profile. Since the \textit{vsini} is probably less than 0.1 km/s for \prox, {\Firstp} based the local line
profile shape for {\ha} on the observed {\prox} line profile since it is unlikely to show rotational broadening. This
profile was tuned to resemble the average {\ha} profile shown in the {\uves} data analysed in \citet{fuhrmeister11}, but
symmetric about the central wavelength. Specifically, {\Firstp} used a Gaussian profile to generate the emission peak
and a second Gaussian with narrow width to represent the central self-absorption.

% as illustrated in Fig. \ref{fig:integregions}.

%\begin{figure}[!htbp]
%\begin{center}
%\includegraphics[scale=0.25]{modelling/images/integregions.png} \\
%\end{center}
%\caption{Example generated model spectrum of \prox, also illustrating the methods for computing the periodicity of
%  spectra.  The centre of the \ha{} line is set at 6563{\AA} for convenience rather than 6562.8\AA{}.  The green lines
%  (from 6561.75{\AA} to 6564.25\AA) show the limits used for calculation of the equivalent width. The blue and red
%  shaded areas (6561.95{\AA} to 6562.75{\AA} and 6563.44{\AA} to 6564.24{\AA} respectively, each 0.8{\AA} wide) show the
%  regions for calculation of the peak ratio.}
%\protect\label{fig:integregions}
%\end{figure}
% Done with 10 plage 80 period 75 deg first spectrum

With {\Firstposs} two-temperature model, in subsequent simulations, {\Firstp} assign either photospheric intensity
or a plage intensity to each pixel. For a pixel containing plage, {\Firstp} thus scale the synthetic {\ha} profile and
for the photosphere with no visible profile (as note above), {\Firstp} use the continuum level. The line profile is
shifted appropriately for the Doppler shift of each pixel in {\Firstposs} model. The model enables {\Firstobj} to place
circular spots of specified radii anywhere on the star. For each viewing angle (or equivalently observation phase),
{\Firstp} calculate the appropriate intensity profiles of all visible pixels (according to position on the line and
centre-to-limb variation) and sum them to obtain our simulated line profile.

A model star with plage regions that rotate into and out of view can thus potentially exhibit variability in the line
shape since the pixels on different parts of the star possess different Doppler velocities. For stars such as \prox,
which possess a \textit{vsini} much less than the instrumental resolution, any distortions in the line
profile due to spots rotating into and out of view may be insignificant or very small. A plage region that rotates into
view may nevertheless have a significant effect on the the equivalent width of the simulated line since {\Firstposs}
local intensity profile for {\ha} possesses a normalised peak intensity of N times the continuum. For stars with
rotational velocity much greater than the instrument resolution, line asymmetries are likely to be much more reliable.

\subsection{Plage distribution and results}
\protect\label{section:plagedist} During the course of experimentation with models, {\Firstp} tried a selection of plage
distributions, ranging from a single large spot on one face to randomly-placed spots of random sizes. However {\Firstp}
found that the variation in equivalent widths from a low spot coverage bore no possible resemblance to that from
observational data, in that just exhibited two extremes of Equivalent Widths and no intermediate values. On the other
hand a coverage of more than about 30\% provided very limited swings in the Equivalent Width compared those observed
from {\harps} and {\uves}. After some experimentation, {\Firstp} settled for randomly distributed plage of random sizes
which filled up to 2.5\% of the surface, towards the high end of the coverage of up to 2.7\% reported in
\citet{guttenbrunner14} in relation to the Sun. \refrevision{Equivalent Widths were calculated for a variety of
  inclinations and starting periods. Peak Ratios were evaluated, but the variations were too small to recover input
  periods. This was also the case for skewness and kurtosis measurements.}

\subsection{Adding in noise and flares}
\protect\label{section:addflares}

Despite the limitations of the simplistic model, it is clear that a good estimate of periodicity, \refrevision{to within
  $\pm$ 0.1 days}, may be obtained from the Equivalent Width method, although the Peak Ratio variations could not be
reproduced and that method reliably applied. These results are for a noiseless set of models and to compare with reality
the performance of the modelling results and the analysis methods in the presence of observational noise and also the
influence of simulated flare events has to be considered.

As a first step in moving to something like actual observational data, {\Firstp} tried adding noise of a given signal to
noise ratio \refrevision{over the whole of the simulated spectra} and observed the effect on the accuracy of the
periodicity measurements for various levels and inclinations. {\FirstP} tried adding Gaussian noise with SNRs from 40
down to 1 in steps of 0.1.  {\FirstP} tried this with all the combinations of inclinations and starting periods tried
before.

It was noticeable that doing this only started to have any significant effect with SNR below 20. Below this level, two
things started to happen, increasingly as the SNR was reduced. Either the error in the recovered period increased,
although not by very much, \refrevision{up to $\pm$ 0.5 days,} alternatively the recovered period was manifestly
  incorrect, giving a clear False Positive such as returning a period of 50 days from a starting period of 80 days.

It was easy to discriminate between these two cases by setting a threshold of 5\% for the difference between the
recovered period and the starting period. If the difference exceeded this, then the period was regarded as incorrectly
recovered, otherwise it was regarded as correctly recovered but with the given error. However in all the cases the
difference was either substantially greater or substantially less than this. It was noticeable that in quite a number of
cases a period close to 116 days was returned as a False Positive, \refrevision{although examination of the window
  function for the observation times did not show this period}.
%, an example of such a case is shown in the right panel of Fig. \ref{fig:noiseresults}.

{\FirstP} also examined the possible effect of flares. {\FirstP} simulated the effect of flares by taking the spectra
which were clipped as having excessive Equivalent Width in Section \ref{section:uvesflares} and adding in the same
proportionate excess over the median Equivalent Width to the model as was found in the observed data. The result was a
poorer performance than with noise alone, but not by much. With just noise, the performance became markedly low with a
SNR of 15 or below but adding flares as described significantly reduced the performance with a SNR of 20 or below.
These values of SNR are much lower than the published values for {\uves} and {\harps}, which are in both cases well over
100.

\section{Discussion and conclusions}
\protect\label{section:discussion}

It is clear that the period of 82.6 $\pm$ 0.1 days given by the photometric results for {\asas} and confirmed by {\hst}
must be the rotation period of \prox, in line with \citet{benedict98} and confirmed by \citet{kiraga07}. There is a
near-zero FAP value and all the routines tried gave exactly the same result. {\FirstP} were not able to obtain as
clear-cut results from spectroscopic methods involving analysis of the {\ha} peak of the {\prox} spectra. The
Equivalent Width and {\ha} Index methods return very similar results but only return the 82.6 day period about 14\% of
the time, never as the strongest period. The Peak Ratio is about four times better.

There is also a strong peak of 106.3 $\pm$ 0.1 days on the {\asas} results and in some of the spectroscopic results and
the modelling, but not seen on the {\hst} results. This period would appear to be a ``beat'' between the rotation period
and an Earth year, which would not affect the {\hst} results, which are far less constrained by the time of year.

It has proved possible to reproduce the 116.6 days of \citet[Table 3]{suarezmascareno15} in both the treatments of
Equivalent Width and {\ha} Index and in some of the other variants of the handling of those, together with periodograms
taken from the skewness and kurtosis measurements, although not often as the strongest peak in the periodogram. Any kind
of selection or binning of the data makes the 116.6-day figure disappear, as does adding in additional {\harps} data
subsequent to 2014. In addition, it was noticed that some of the modelling results which failed to give the expected
period (see Section \ref{section:addflares}) also gave periods close to 116.6 days from the same observation times as in
the {\harps} data. From this analysis, {\Firstp} would have to discount this as a false positive, most probably an
artefact of these observation times.

Limiting the portion of the spectrum to just the {\ha} line of {\prox}, even with the instrumental stability of
{\harps}, was proven to be less useful than {\asas} ground-based photometry for the recovery of period. A future line of
investigation which might be worth considering is that of combining fluxes from various magnetic/activity sensitive
lines in various spectral orders to re-evaluate the Composite Spectral Index referred to in \citet{hall99} and
\citet{hall00}.

{\FirstP} were able to reproduce the variations in Equivalent Width seen in {\prox} using the DoTS model and show that
the recovery of the rotation period has validity. However, even with extensive experimentation, including relatively
extreme values for the various parameters for limb-darkening and contrast or extreme distributions of plage, {\Firstp}
could not model the observed variations in Peak Ratio found either in the {\harps} or {\uves} data. It is clear that the
variations in the two \horn s that are purely due to Doppler shift from the rotational velocity are not large; with a
radius of 0.141 Solar \citep{demory09} and assuming a period of the order of 80 days the rotational velocity is at most
90 m/s yielding a Doppler shift of at most 0.003{\AA} in the {\ha} line between the extremes of the disk and the centre,
far too low to reproduce the variations in the \horn s in the {\ha} line profile as illustrated in
Fig. \ref{fig:harpsfirstha}, for which the Peak Ratios were calculated in Section \ref{section:linemeas} as 0.994 $ \pm
$ 0.017, whereas the best standard deviation on a Peak Ratio close to 1.0 which could be obtained from the models was
$2{\times}10^{-5}$ or $2{\times}10^{-4}$ for very extreme plage distributions. This was not surprising due to the lack of
Doppler broadening of the line profile.

It is clear that the 2D model of static plage and spots supported by DoTS cannot reproduce the observed variations in
the Peak Ratios. Likewise it cannot reproduce the range of phenomena which adversely affects obtaining periodicity from
the Equivalent Widths. \refrevision{{\FirstP} did consider the possibility of differential rotation affecting the
  spectroscopic results, but discounted this in the light of finding no evidence in the available datasets and
  \citet{barnes05}, which argues that differential rotation decreases with decreasing stellar mass.}

This points to the need for a 3D model including vertical processes to properly understand the
behaviour of \prox. In \citet{mohanty02} and \citet{mohanty03}, where the activity of late \rdwarf s is found to be less
closely tied to the rotation period than for earlier type stars, the authors propose a ``turbulent dynamo'' as the
source of the activity, for which a 3D model is required. In consideration of this, {\Firstp} note the success of 3D MHD
simulation for the Sun in \citet{leenaarts12} and also the work on seismic shock waves such as in
\citet{donea06}. Similar conclusions are reached by \citet{rauscher06} as an explanation for H, K and Ca line asymmetry,
where-the authors suggest that this is caused by slowly-decelerating motion toward the observer which does not fall back
ballistically.
%
%\section{Conclusions}
%\protect\label{section:conclusions}
%
%{\FirstP} found that the methods involving study of {\ha} were not sufficiently reliable on their own, at least as far
%as {\prox} is concerned, for them to be of use on their own for that star over and above photometric measurements,
%although the Peak Ratio method introduced here has promise. Nonetheless, increasing confidence in previous
%calculations of the rotation period of {\prox} of between 82.5 and 83.0 days emerged during this process and a recent
%derivation of 116.6 days using the {\ha} Index was revealed as a false positive.
%
%{\FirstP} have also attempted to model \prox, with some success in reproducing the Equivalent Width variations and
%observing the effects of noise and flares on the accuracy of the recovered periods. The modelling does clearly need
%improvement to reflect and aid understanding of the both the observed variations in the Peak Ratio and the other factors
%which reduce the effectiveness of the Equivalent Width and similar measurements in the observational data.

\section{Acknowledgements}

The {\harps} data was obtained from HARPS public database at the European Southern Observatory (ESO), programs
072.C-0488(E), 082.C-0718(B), 183.C-0437(A) and  191.C-0505(A).

{\FirstP} also wish to thank Birgit Fuhrmeister and Lalitha Sairam for providing {\Firstobj} with some additional data
referred to in \citet{fuhrmeister11}.

{\FirstP} acknowledge the value of the Vienna Atomic Line Database (VALD) for spectral line data and the {\asas} for
additional observations of {\prox}.

All the figures were produced \refrevision{via Python programs developed by the authors using the {\matplot} libary}, which is
associated with the {\scipy} library \citep{jones01}.

{\FirstP} are grateful to the anonymous referee for his or her kind assistance in recommending significant improvements
to prepare this paper for publication.

%\begin{appendices}
%\input{appmodparams/modparams.tex}
%\input{appmdwarfs/mdwarftable.tex}
%\input{appcompar/mdwarfcompar.tex}
%\end{appendices}

\bibliographystyle{aa}
\bibpunct{(}{)}{;}{a}{}{,} % to follow the A&A style
%\bibpunct{(}{)}{;}{s}{,}{,}
\bibliography{bibrefs}

\protect\label{lastpage}
\end{document}